\documentclass[epj,draft]{svjour}
\usepackage[]{psfig}

\begin{document}
\title{Force measurements in positive unipolar wire-to-plane corona discharges in air}

\author{Victor de Haan}

\institute{BonPhysics B.V., Laan van Heemstede 38, 3297 AJ
Puttershoek, The Netherlands, \email{victor@bonphysics.nl}}

\date{Received: date / Revised version: date}
%
\abstract{ Measurements of force generated by a positive unipolar
wire-to-plane corona discharge in air are compared with numerical
simulations. The generated force does not depend on the ion or
electron mobilities, preventing the influence of uncertainty and
variation of these parameters. A method is described to simulate the
voltage and charge distribution for a wire-to-plane set-up. This
method enables the determination of the transition between unipolar
and bipolar discharges. In the experimental set-up breakdown
electric field of air reduces with increasing discharge current.
} 

\maketitle

\section{Introduction}
Atmospheric pressure corona discharges in air are important for use
as electrostatic precipitators, ozone generators~\cite{Chen},
thunderclouds~\cite{Aleksandrov}, mass transport~\cite{Sigmond} and
so on. An upcoming field is the use of corona discharge to generate
electric winds for cooling purposes~\cite{Schlitz}. The physics of
corona discharge however is not fully understood. Peek~\cite{Peek}
was one of the first to study the phenomenon in a systematic way. He
found the onset breakdown electric field at the wire surface of a
wire-cylinder discharge corresponds to
\begin{equation} \label{peekeq}
E_a = f E_o \delta \left( 1 + \sqrt{ \frac{a_o}{ a \delta } }
\right)  \  \ ,
\end{equation}
where $f$ is an irregularity factor (for polished copper wire $f
\approx 1$), $E_o$ the breakdown electric field of air equal to 31
kV/cm, $\delta $ the relative density of the air compared to the
density of air at 25 $^o$C and 1013 kPa, $a$ the wire radius and
$a_o$ a constant equal to 0.0949 cm. In this paper the outer edge of
the corona is defined as the distance to the center of the wire,
$r_c$ where the electric field equals the breakdown electric field.
In the corona between wire surface and outer corona edge it is
assumed that space charge has negligible influence on the electric
field. This is a valid assumption as close to the wire the electric
field gradient is already very large due to geometry. Then,
according to equation~(\ref{peekeq}) the corona size is given by
\begin{equation} \label{peekrc}
r_c = a + \sqrt{ \frac{a_o a}{ \delta} }  \  \ .
\end{equation}
This can be calculated by using equation~(\ref{EWire}) with
$\alpha(x)=1$. The second term at the right hand side of this
equation represents the ionisation sheath. In this sheath the
discharge is carried by electrons, positive and negative
ions~\cite{Chen2}. Outside this sheath the discharge is mainly
unipolar. A recent overview of possible electrical breakdown
criteria has been given by Lowke~\cite{Lowke}. He couples $a_o$ to
the discharge parameters of air
\begin{equation} \label{lowkea0}
a_o = \frac{\ln(Q)}{(E_o/N_o)^2N_oB}  \  \ ,
\end{equation}
where $Q$ is the number of ionisations to occur to sustain
breakdown, $N_o$ is the air number density at 25 $^o$C and 1013 kPa
and $B$ is 1.35$\times 10^{16}$~V$^{-2}$m$^{-2}$ \footnote{Lowke
uses $B$=2.08$\times 10^{16}$~V$^{-2}$m$^{-2}$, but he takes 25
kV/cm for the breakdown electric field of air.}. $Q$ depends on the
corona discharge mechanism. He suggests a possible mechanism for the
onset corona discharge but does not take into account the effect of
space charge in the air surrounding the corona. Normally this is not
a problem as measurements of the breakdown electric field of air are
mostly done for relatively small space charges at relatively small
discharge currents. However, for the above-mentioned applications
the breakdown electric field in the presence of a discharge current
is important. Additional problems are the relative uncertainties of
the mobility of the involved ions and electrons. In the following it
is shown that the force generated by a wire-to-plane corona
discharge is independent of charge carrier mobilities. This force
only depends on the applied potential difference between wire and
plane and the geometry of the discharge. Force measurements were
performed to determine the influence of space charge on the
breakdown electric field of air at atmospheric pressure and to
eliminate the influence of the charge carrier mobilities. In the
following this method is described and measurements for several
geometries are presented and discussed.

\section{Outline of the method}
The method is based on solving the Poisson equation for certain
geometry combined with the continuity equations for ion and electron
flow. Solving Poisson equation gives the potential distribution, $U$
and the continuity equations give the space charge distribution,
$\rho$. In air (the relative permittivity, $\epsilon_r$ is
approximated by 1) the Poisson equation is
\begin{equation} \label{Poisson}
\nabla^2 U = -\frac{\rho}{\epsilon_o} \ \ ,
\end{equation}
where $\epsilon_o$ is the permittivity of free space. For each
charge species $i$, being positive or negative ions or electrons,
the stationary continuity equation is
\begin{equation} \label{Continuity_i}
\nabla.\vec{J_i} = 0 \ \ ,
\end{equation}
where $\vec{J_i}$ is the current density of charge species $i$. It
is assumed that no transformation from one species into another
occurs. If the mobility, $\mu_i$ of the species does not depend on
either space charge nor electric field, $\vec{E}=-\nabla U$ it is
given by
\begin{equation}
\vec{J_i} = \mu_i | \rho_i | \vec{E} \ \ .
\end{equation}
where $\rho_i$ is the charge density of species $i$. Hence,
equation~(\ref{Continuity_i}) reduces to
\begin{equation} \label{Cont2}
 \rho_i \nabla.\vec{E} + \vec{E}.\nabla \rho_i = 0\ \ .
\end{equation}
Adding all equations~(\ref{Cont2}) results in the continuity
equation for the total space charge
\begin{equation} \label{Cont_tot}
 \rho \nabla.\vec{E} + \vec{E}.\nabla \rho = 0\ \ .
\end{equation}
The force exerted by the electric field on a point particle with
charge $q$ is by definition $\vec{F}=q\vec{E}$. This can be extended
to find the force exerted by the electric field on a charge density
distribution in a volume $V$:
\begin{equation} \label{Force}
\vec{F} = \int \rho \vec{E} dV \ \ .
\end{equation}
Neither equation~(\ref{Poisson}) nor~(\ref{Cont_tot}) contains any
reference to the mobility of the involved ions or electrons, other
then that they are constant. The simultaneous solution of these
equations results in voltage and charge density distributions
independent of these parameters. Hence according to
equation~(\ref{Force}) the force exerted by the charge distribution
does not depend on the mobilities. Further, this force is uniquely
determined by the applied voltage difference between wire and plane
and the geometry. The electric current, $\vec{I}$ through some flat
area $A$ is given by
\begin{equation} \label{Current}
\vec{I} = \sum_i \int \vec{J_i}.\vec{n} \  dA = \sum_i \int  \mu_i
|\rho_i| \vec{E}.\vec{n} \ dA \ \ ,
\end{equation}
where $\vec{n}$ is the normal to area $A$. The current consists of
contributions from all charge species. Hence, it is much more
difficult to calculate accurately than the force. Further, if the
total charge distribution consists of positive and negative species,
the current increases as indicated by the absolute sign.

\section{Experimental set-up}
\begin{figure}
\begin{picture}(100,180)
\put(10,0){\psfig{figure=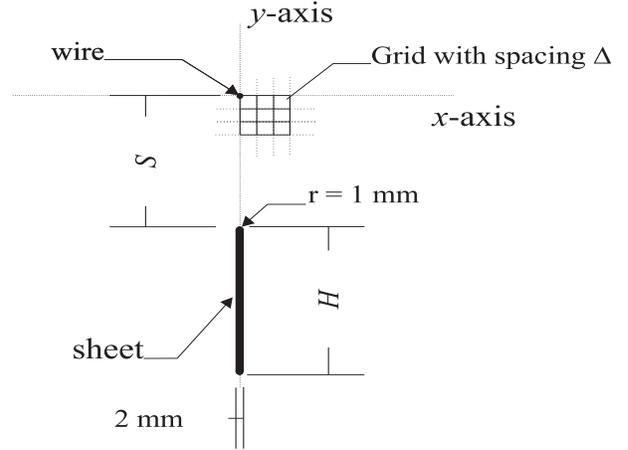,height=60mm,width=80mm}}
\end{picture}
\caption{\label{setup}Sketch of a cross section of the wire-to-plane
geometry}
\end{figure}

The detection of force exerted by a space charge distribution needs
equation~(\ref{Force}) to produce a result different from 0. This
can be achieved by an asymmetrical corona discharge, like a
wire-to-plane discharge. Normal the wire is put above and parallel
to the plane. Then, the electric wind generated by the space charge
will push against the plane and reduce the measurable
force~\cite{Bequin}. To overcome this problem, the plane is reduced
to a sheet and the wire is put parallel to the sheet at a distance
$S$. The wire is positioned in the same plane as the sheet. $H$ is
the height of the sheet (4.0 cm in the experiments). This
configuration is shown in figure~\ref{setup}.
Sheet thickness was 2 mm, far less than either $S$ or $H$. The upper
and lower end of the sheet were smoothly rounded with a radius of 1
mm to as much as possible avoid production of opposite signed
charges. The length of both wire and sheet was 53.5~cm. In the
experiment $S$ was varied between 2.0 and 6.0~cm with steps of
1.0~cm. Wires of different radius $a$ were used: 0.040, 0.0635,
0.075, 0.10, 0.1575 and 0.25~mm. Measurements were performed at
atmospheric conditions: 20~$\pm$~1~$^o$C and 1013~$\pm$~10 kPa.
Relative humidity was between 0.5 and 0.8. The force was measured by
the reduction of weight of the set-up with a lever. One arm of the
lever was connected with light non-conducting wires attached to the
set-up. The length of these wires was about 50 cm ensuring the
influence of the surroundings was negligible. The other arm was
rigidly connected to a balance with a range of 100 gram and an
accuracy of 0.1 gram. The ratio of the length of the arms of the
lever was 1:10. This lever enabled force measurements between 0 and
100 mN with an accuracy of 0.1 mN. The measurements were done with
increasing and decreasing voltages to check for systematic
deviations. A DC high voltage between 0 and 30 kV was applied to the
wire while the sheet was grounded. An ampere meter monitored the
current supplied by the high voltage power supply.

\section{Simulation}
\begin{figure}
\begin{picture}(100,200)
\put(0,0){\psfig{figure=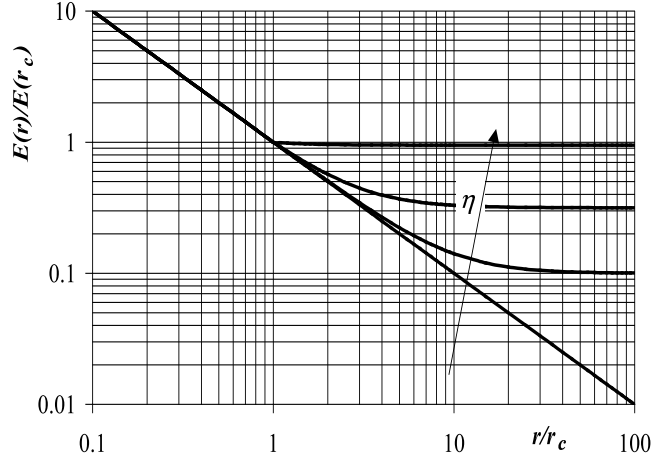,height=60mm,width=85mm}}
\end{picture}
\caption{\label{efield}Ratio of electric field,
 $E(r)$ to electric field at corona edge, $E(r_c)$ as function of distance to center of wire, $r$ for
 $\eta = 0; 0.01; 0.1$ and $0.9$. The arrow indicates increasing values of $\eta$.}
\end{figure}

\begin{figure}
\begin{picture}(100,200)
\put(0,0){\psfig{figure=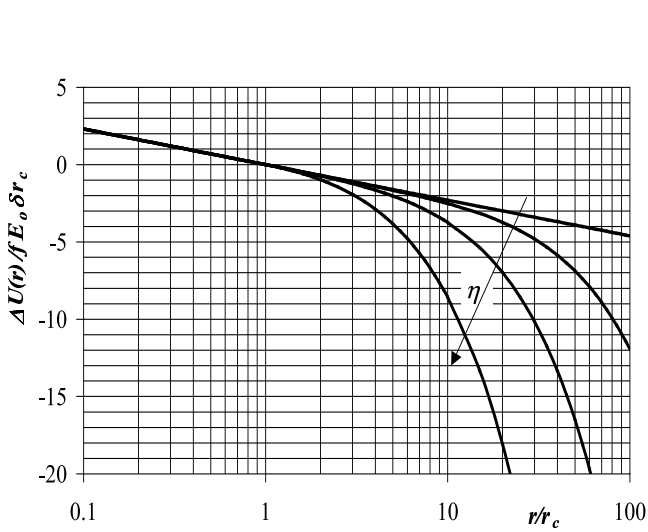,height=60mm,width=85mm}}
\end{picture}
\caption{\label{VoltageDrop}Normalized voltage drop $\Delta U(r)/f
E_o \delta r_c $ as function of distance to wire center, $r$ for
$\eta = 0; 0.01; 0.1$ and $0.9$. The arrow indicates increasing
values of $\eta$.}
\end{figure}

To simulate the electric field a model was developed
following~\cite{McDonald}. First, the problem is reduced to 2
dimensions by assuming an infinitely long wire and sheet. Second,
the problem is made periodic in both $x$- and $y$-directions with a
period of $2W_x$ and $W_y$ respectively. $2W_x$ respectively $W_y$
being the distance in $x$-respectively $y$-direction between two
adjacent set-ups. $x = W_x$ and the $x = 0$ are symmetry axis. If
$W_x$ and $W_y$ are large enough the results can be applied to a
single set-up. Here, $W_x$ and $W_y$ are chosen optimizing the
computation time and required accuracy. Third, the problem is
discretized by assuming a square grid with spacing $\Delta$. The
number of grid lines in the $x$-direction, $n$ equals
$W_x/\Delta+1$, the number in the $y$-direction, $m$ equals
$W_y/\Delta$. Fourth, the appropriate boundary conditions are
defined and the discretized version of equations~(\ref{Poisson})
and~(\ref{Cont_tot}) are solved sequentially until a stable solution
is found. For the solution of the discretized Poisson equation a
Poisson solver was used based on the linear conjugate gradients
method~\cite{CG}. For the solution of the descritized continuity
equation a non-linear conjugate gradient method was
used~\cite{NLCG}. The appropriate boundary conditions are not
evident. At the sheet the potential was simply set to 0. At the wire
position however, the wire radius had to be taken into account. As
the wire radius is very small compared with the grid size a method
had to be devised to take it into account. The method as described
in~\cite{McDonald} was used and extended to incorporate unipolar
space charge around the wire. Around the wire up to the first
adjacent grid point cylindrical geometry is assumed. This is
justified as $\Delta \ll S$. The space charge density at $r=r_c$ is
taken to be $\rho_c$. Then, the electric field and charge density
near the wire are given by
\begin{equation} \label{EWire}
E(r) = f E_o \delta \frac{\alpha(r/r_c)}{r/r_c}  \ \  ,
\end{equation}
\begin{equation} \label{RhoWire}
\rho(r) = \frac{\rho_c}{ \alpha(r/r_c) } \ \  ,
\end{equation}
where $\alpha(x) = \sqrt{1+\eta(x^2-1)}$ if $x \geq 1$ and
$\alpha(x) = 1$ if $x < 1$ and $\eta=\rho_cr_c/fE_o\delta\epsilon_o$
is a measure for the total space charge at the outside of the corona
($r~=~r_c$). For a unipolar discharge (outside of the corona) $\eta$
is proportional to the discharge current, $I$
\begin{equation} \label{EtafromI}
\eta = \frac{I/L}{2\pi\mu \epsilon_o(f E_o \delta)^2}  \ \  ,
\end{equation}
where $\mu$ is the mobility of the unipolar charge carriers. Note
that $\eta$ is determined by the experimental conditions only. A
graph of the electric field as function of $r$ for several $\eta$ is
shown in figure~\ref{efield}.
The theoretical maximum value for $\eta$ is 1. For $\eta=1$ the
electric field in the whole gap equals the breakdown electric field
and there is no way to prevent spark discharge. For $r/r_c < 1$ the
lines coincide. Inside the corona the electric field gradient due to
the space charge density is negligible compared to the geometric
gradient. For all experiments $\eta < 0.05$, so the influence of the
space charge on the electric field is negligible near the wire up to
$r=5r_c$. Hence, the exact shape of the streamers in the corona area
is not important for the results presented here. The voltage drop
from wire to outer edge of the corona is given by
\begin{equation} \label{DUCorona}
 \Delta U_c= -f E_o \delta r_c \ln{\frac{r_c}{a}} \ \ ,
\end{equation}
independent of $\eta$. The voltage drop from outer corona edge to a
radius of $r > r_c$ is given by
\begin{equation} \label{DUWire}
 \Delta U(r) = -f E_o \delta r_c \left( \alpha(\frac{r}{r_c})-1+
\chi(\frac{r}{r_c}) \sqrt{1-\eta}  \right) \ \ ,
\end{equation}
where
\begin{equation}
\chi(x) = \ln \{\frac{\sqrt{1-\eta}+1}{\sqrt{1-\eta}+\alpha(x)}x \}
\ \ .
\end{equation}
This function is shown in figure~\ref{VoltageDrop} for different
values of $\eta$.
Note the positive values for the voltage drop as $r/r_c < 1$. Here,
$\Delta U(r) = -f E_o \delta r_c \ln (r/r_c)$, again independent of
$\eta$. For the boundary conditions at the wire the electric field
at all grid positions adjacent to the wire is made equal to
$E(\Delta )$ and the space charge density is set to $\rho(\Delta )$.

\section{Results and discussion}
\begin{figure}[htbp]
\begin{picture}(100,200)
\put(0,15){\psfig{figure=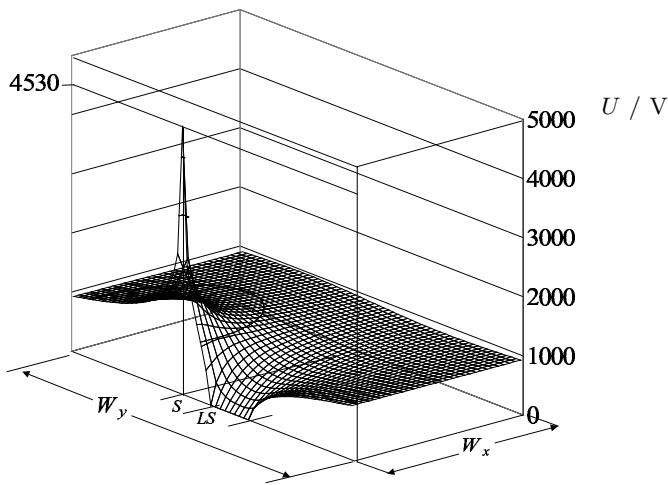,height=180pt,width=215pt}}
\put(225,155){$U$ / V}
\end{picture}
\caption{\label{VolDistrib}Voltage distribution, $U$ at the onset
voltage for $S$ = 3.0~cm, $LS$ = 4.0~cm and $a$ = 0.040~mm. Grid
size was 5.0~mm; $W_x$ = 15.0~cm; $W_y$ = 30.0~cm. Wire onset
voltage = 5.82~kV. }
\end{figure}

\begin{figure}
\begin{picture}(100,200)
\put(0,0){\psfig{figure=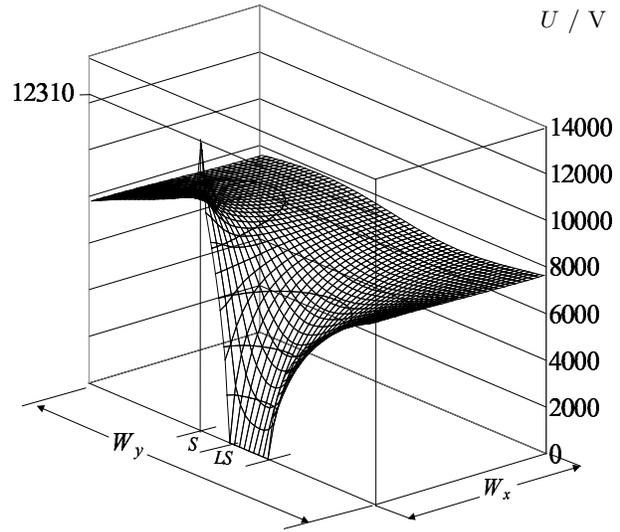,height=70mm,width=80mm}}
\put(200,190){$U$ / V}
\end{picture}
\caption{\label{DistribIV}Voltage distribution, $U$ for $\rho_c$ =
128~$\mu $Cm$^{-3}$ and same dimensions as in
figure~\ref{VolDistrib}. Wire voltage = 13.6 kV. }
\end{figure}

\begin{figure}
\begin{picture}(100,200)
\put(0,0){\psfig{figure=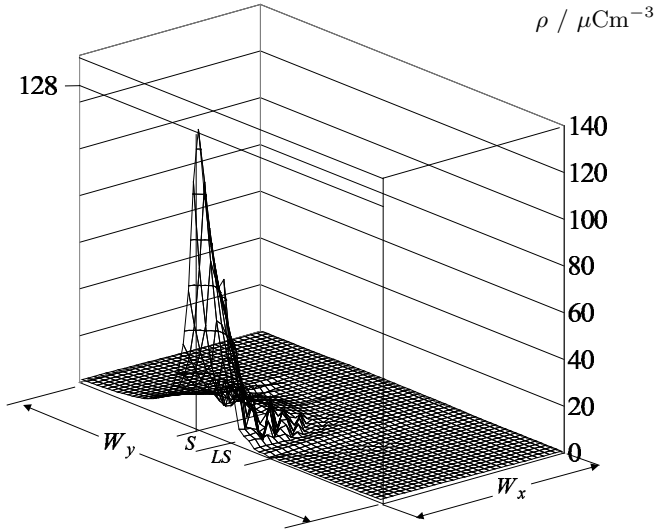,height=70mm,width=80mm}}
\put(200,190){$\rho$ / $\mu $Cm$^{-3}$}
\end{picture}
\caption{\label{DistribIR}Charge density distribution, $\rho$
 for $\rho_c$ = 128~$\mu $Cm$^{-3}$ and same dimensions as in figure~\ref{VolDistrib}. Wire voltage is 13.6~kV. }
\end{figure}

\begin{figure}
\begin{picture}(100,215)
\put(25,5){\put(0,20){\psfig{figure=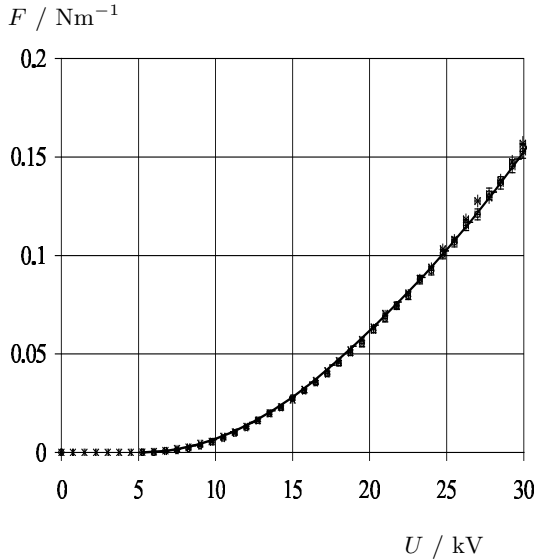,height=60mm,width=70mm}}
\put(0,200){$F$ / Nm$^{-1}$} \put(150,0){$U$ / kV}}
\end{picture}
\caption{\label{FUF} Exerted force, $F$ as function of wire voltage,
$U$ for same dimensions as figure~\ref{VolDistrib}. The symbols with
error bars represent the measurements for increasing (crosses) and
decreasing (circles) voltage. The lines represent the simulated
results. }
\end{figure}

\begin{figure}
\begin{picture}(100,200)
\put(20,0){
\put(0,30){\psfig{figure=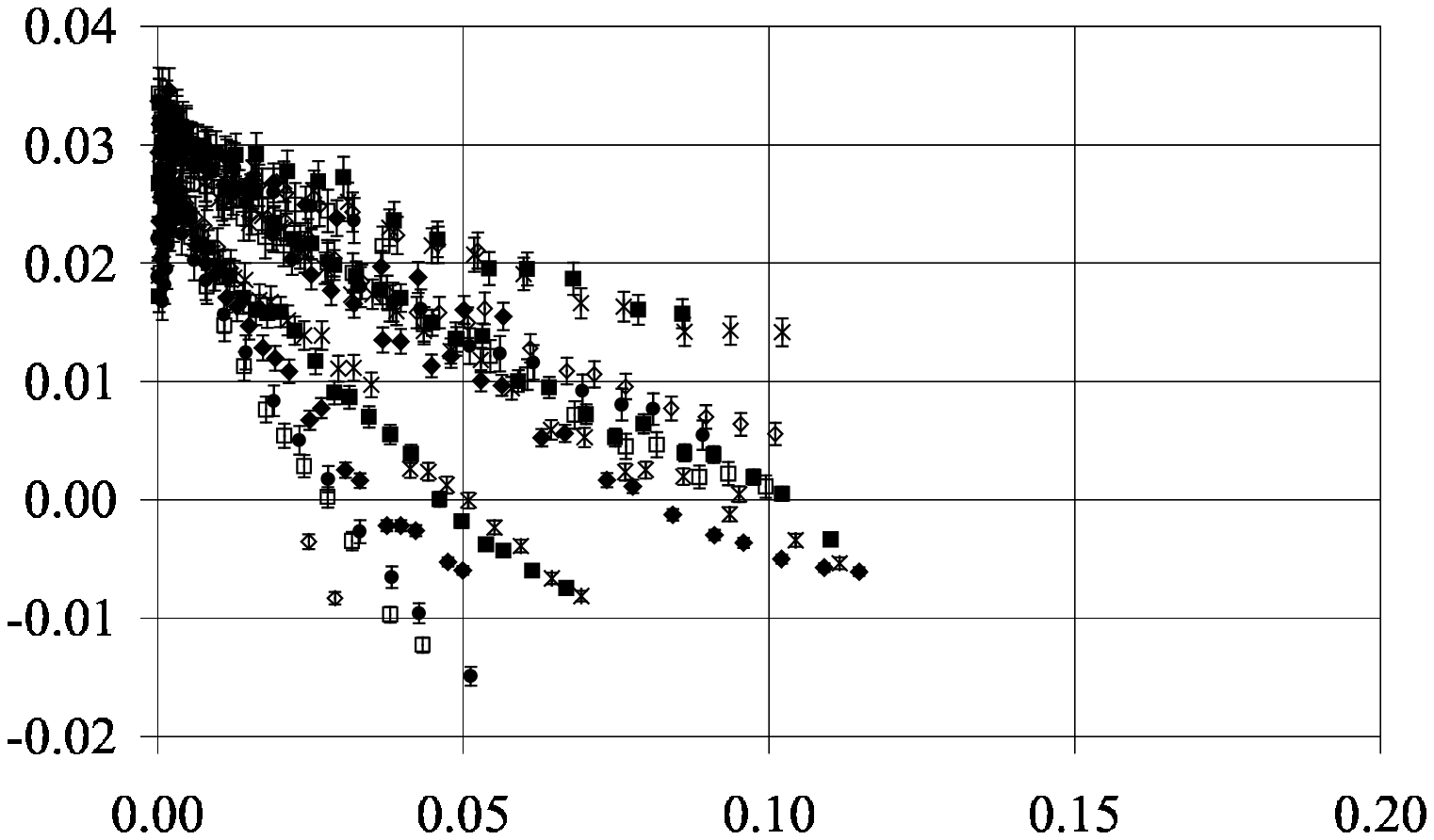,height=60mm,width=70mm}}
\put(0,30){\put(150,-10){$F$ / Nm$^{-1}$}
\put(-20,140){$\sqrt{a_o}$} \put(-20,120){m$^{0.5}$}
\put(120,100){$S$ = 2.0 cm} \put(95,90){\thicklines \line(0,1){30}}
\put(120,50){$S$ = 4.0 cm} \put(80,70){\thicklines \line(0,1){30}}
\put(70,30){$S$ = 6.0 cm} \put(45,35){\thicklines \line(0,1){70}}}}
\end{picture}
\caption{\label{sqrta01}Calculated $\sqrt{a_o}$ as function of
measured force $F$ for measured geometries with $S$ = 2.0; 4.0 and
6.0 cm. Closed diamonds: $a$=0.040 mm; closed squares:
$a$=0.0635~mm; stars: $a$=0.075~mm; open diamonds: $a$=0.10~mm; open
squares: $a$=0.1575~mm and dots: $a$=0.25~mm. Error bars are
estimates of accuracy of simulation. The vertical lines indicate the
maximum force for which the discharge is unipolar.}
\end{figure}

\begin{figure}
\begin{picture}(100,200)
\put(20,0){\put(0,30){\psfig{figure=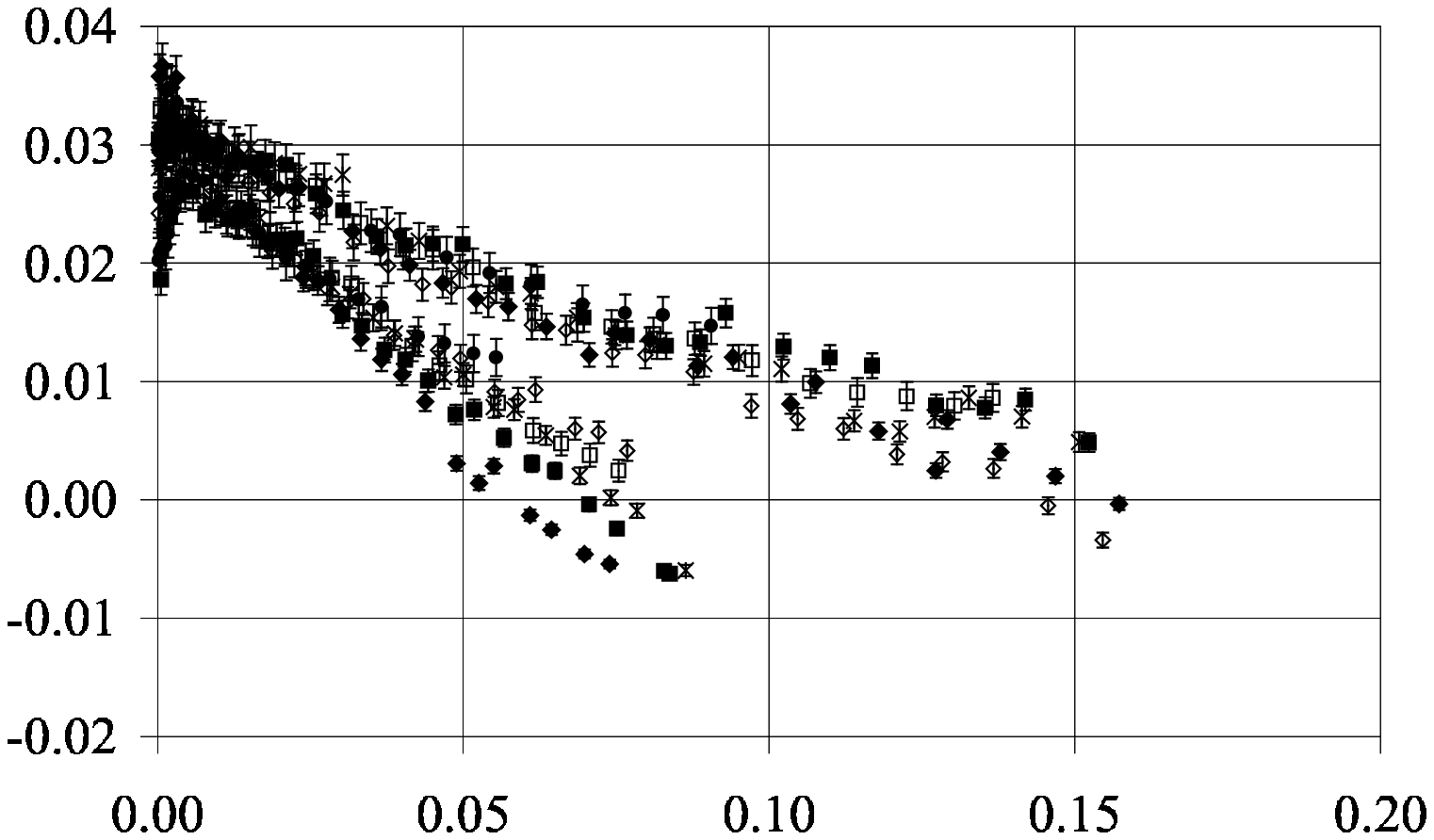,height=60mm,width=70mm}}
\put(0,30){ \put(150,-10){$F$ / Nm$^{-1}$}
\put(-20,140){$\sqrt{a_o}$} \put(-20,120){m$^{0.5}$} \put(90,40){$S$
= 5.0 cm} \put(60,67){\thicklines \line(0,1){40}} \put(130,100){$S$
= 3.0 cm} \put(80,90){\thicklines \line(0,1){35} } }}
\end{picture}
\caption{\label{sqrta02}Calculated $\sqrt{a_o}$ as function of
measured force $F$ for measured geometries with $S$ = 3.0 and 5.0
cm. See figure~\ref{sqrta01} for details.}
\end{figure}

The wire voltage was calculated by adding $\Delta U(\Delta )+\Delta
U_c$ to the average of the voltage at the wire adjacent grid points.
From the simulations the spread in the voltages at these grid point
was only a few percent. This indicates the accuracy of the
assumption of cylindrical geometry near the wire up to the first
grid point. $f$ was determined by fitting the onset breakdown
electric field to equation~(\ref{peekeq}). For the onset $r_c$ was
calculated with equation~(\ref{peekrc}). As an example, the onset
voltage distribution for $S$~=~3.0~cm and $a$~=~0.040~mm is shown in
figure~\ref{VolDistrib}.
The position of the wire and sheet are evident. Notice the steep
rice at the wire position. The wire onset voltage was 5.82~kV. In
this case the corona radius according to equation~(\ref{peekrc}) is
0.235~mm. The corona voltage drop according to
equation~(\ref{DUCorona}) was ($f$~=~1) 1290~V. In the simulation
the voltage at the wire position equals the voltage at the outside
of the corona. Hence, in the simulation the voltage at the wire
position was 4.53~kV. The fitted irregularity factors are shown in
table~\ref{tab1}.
\begin{table*} \begin{center}
\begin{tabular}{|l|l|l|l|l|l|l|l|l|l|}
\hline
 $S$ & $W_y$ & $W_x$ & $\Delta$ & $a$ / mm  &  &  &  &  &    \\
  cm & cm    & cm    & mm       & 0.040     & 0.0635 & 0.075 & 0.10 & 0.1575 & 0.25       \\
\hline
 2.0 & 20.0 & 10.0 & 2.0 & 0.88 & 0.89 & 0.87 & 0.87 & 0.86 & 0.85    \\
 3.0 & 33.0 & 12.0 & 3.0 & 0.90 & 0.90 & 0.83 & 0.83 & 0.90 & 0.85    \\
 4.0 & 40.0 & 20.0 & 4.0 & 0.85 & 0.78 & 0.88 & 0.88 & 0.91 & 0.85    \\
 5.0 & 50.0 & 25.0 & 5.0 & 0.80 & 0.80 & 0.85 & 0.85 & 0.81 & 0.95    \\
 6.0 & 60.0 & 30.0 & 6.0 & 0.80 & 0.80 & 0.80 & 0.80 & 0.90 & 0.95    \\
\hline
 average & & & & 0.86(4) & 0.84(5) & 0.84(5) & 0.86(2) & 0.86(2) & 0.88(2)      \\
\hline \end{tabular} \caption{Fitted irregularity factor $f$ and
simulation parameters for all geometries. $f$ was fitted to match
the measured onset electric field at the wire position with
equation~\ref{peekeq}. } \label{tab1}
\end{center}
\end{table*}
The onset voltages varied between 5 and 12~kV. The average
irregularity factor is 0.86, this indicates that the used value of
$E_o$ is slightly too high. Using a value of 27~kV/cm instead of
31~kV/cm, increases the average irregularity factor to 0.99. The
small spread in the fitted value of the irregularity factor is
evidence for the correctness of the performed simulation. Using
these irregularity factors, at every measured voltage and geometry
the simulated force was fitted to the measured one by adjusting
$r_c$ and $\rho_c$. Both parameters are needed to determine the
force at a certain voltage and geometry. All other parameters were
taken from the experimental conditions. Figure~\ref{DistribIV} shows
an example of a voltage distribution for
$\rho_c$~=~128~$\mu~$Cm$^{-3}$ (which corresponds to a unipolar
discharge current of 0.1~mA/m, if
$\mu$~=~2.0~cm$^2$V$^{-1}$s$^{-1}$) with the same geometry as in
figure~\ref{VolDistrib}.
Figure~\ref{DistribIR} shows an example of the corresponding space
charge density distribution. Notice the small oscillations in the
current density. This is due to the limited accuracy of the
calculations and the accuracy of discretization. It is assumed they
do not significantly change the results presented here.
Simulations and measurements of the exerted force for the
above-mentioned geometry as a function of applied wire voltage are
shown in figure~\ref{FUF}.
\begin{figure}
\begin{picture}(100,215)
\put(25,5){\put(0,20){\psfig{figure=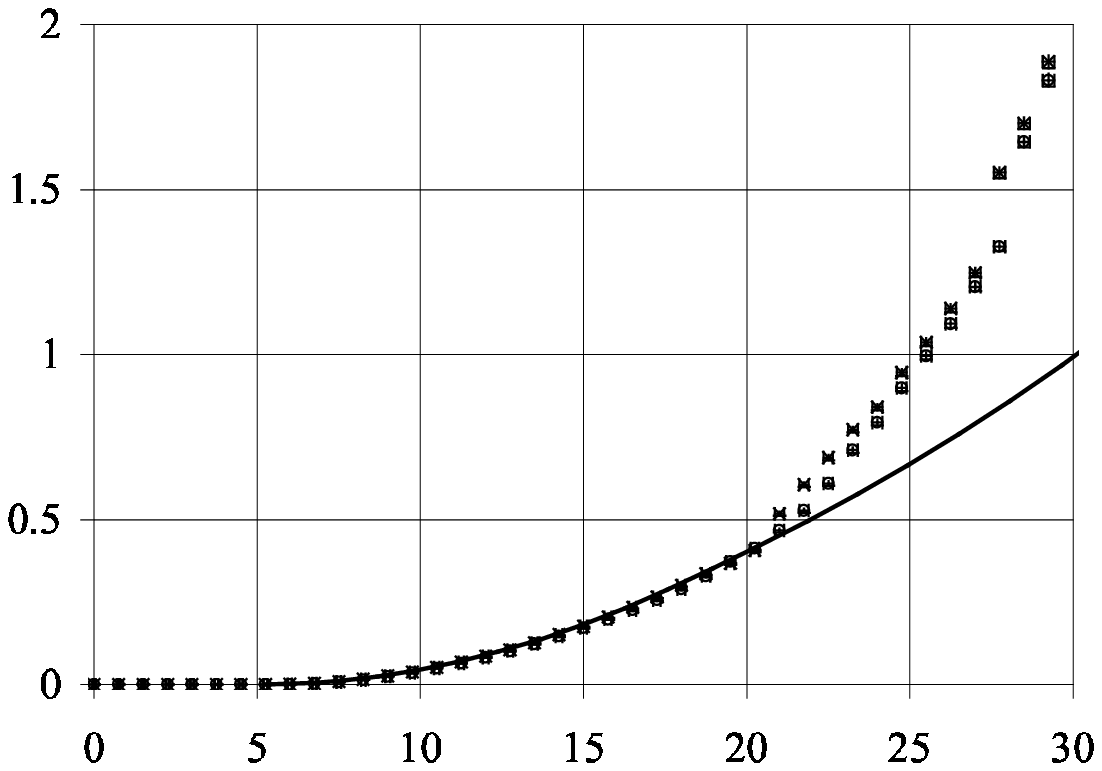,height=60mm,width=70mm}}
\put(0,200){$I$ / mAm$^{-1}$} \put(150,0){$U$ / kV}}
\end{picture}
\caption{\label{FUD} Discharge current, $I$ as function of wire
voltage, $U$ for same dimensions as figure~\ref{VolDistrib}. The
symbols with error bars represent the measurements for increasing
(crosses) and decreasing (circles)
 voltage. The lines represent the simulated results. }
\end{figure}
The symbols with error bars represent the measurements for
increasing and decreasing voltage. The lines represent the simulated
results. The fitted force matches the measurement accurately. The
discharge current is calculated using the voltage and charge
distributions and equation~(\ref{Current}) applied for a unipolar
current with a mobility of 2.0~cm$^2$V$^{-1}$s$^{-1}$. This is show
in figure~\ref{FUD}. For small voltages the deviation with the
measurements are within the accuracy of the measurements. Note that
both force and discharge current measurements are accurately
calculated with the same space charge density distribution. For the
force this is a result of the fit. For the discharge current this is
not evident, as the currents were not used in the fits. Force is a
volume and current an area integral over a charge and electric field
distribution corresponding to equation~\ref{Force} and~\ref{Current}
respectively. Hence, good agreement between the measured and
simulated discharge current is a further indication of the validity
of this method. For larger voltage the measured discharge current
grows faster than the simulated one. Here, the current measurements
with increasing and decreasing voltage starts to deviate slightly.
This is caused by a change of the discharge from unipolar to
bipolar. The charges of the other sign reduce the force but enhance
the current. From the fitted $r_c$ the corresponding values of
$\sqrt{a_o}$ are calculated using equation~(\ref{peekrc}). They are
shown in figures~\ref{sqrta01} and~\ref{sqrta02}.
The vertical lines in the graphs show the maximum force for which
the simulated current matched the measured one. This is the maximum
force for a unipolar discharge in the specified geometry.
$\sqrt{a_o}$ decreases from the onset value of 0.0308 m$^{0.5}$ more
or less linear with a slope dependent on the geometry. The slope is
proportional to $S$. For different wire diameters the slope is
approximately the same. The reduction of $\sqrt{a_o}$ corresponds to
a reduction of the electric field at the wire position. In
literature this is attributed to a reduction of the breakdown
electric field of air dependent on the applied over
voltage~\cite{Jaiswal,Aboelsaad}. The physics behind this are not
yet understood. The reduction of $\sqrt{a_o}$ to $0$ corresponds,
according to equation~(\ref{lowkea0}), to a reduction of $Q$ to 1.
The number of needed ionisations reduces to 1. The slope's
independence of $a$ and its proportionality to $S$ suggest $Q$
depends on the total amount of charge present in the discharge and
not on local conditions around the wire. Lowke proposes the
breakdown of air is caused by ions produced by interaction of
metastable molecules. The effect presented here supports his
conclusion, as the number of metastable molecules present in the air
depends on the charge density. The measurements presented here are
not accurate enough to determine the mechanism of this dependence.
For larger forces, where the discharge becomes bipolar, $\sqrt{a_o}$
becomes negative, indicating the limit of equation~(\ref{Cont_tot}).
This equation assumes no transformation of one charge species into
another. This is further supported by the observed trend that for
larger interaction times of the charge species (i.e. for large
traveling distance of the charges) $\sqrt{a_o}$ becomes negative for
smaller forces.

\section{Conclusions}
Force generated by an asymmetrical wire-to-plane discharge can be
measured and accurately simulated. Comparison of measurements of
both force-voltage and current-voltage characteristics with finite
differences simulations show good agreement for the unipolar region.
These measurements can be used to determine the transition voltage
of an unipolar into a bipolar discharge. Further, it is possible to
determine the breakdown electric field of air dependent on the
discharge conditions. The presented measurements are in qualitative
agreement with the breakdown mechanism as proposed by Lowke, but
measurements are not accurate enough to determine the dependence of
this mechanism on the space charge density distribution.

\begin{acknowledgement}The author wishes to thank Mrs. G. Gravendeel-van Ingen for
a valuable contribution to the experimental set-up.
\end{acknowledgement}


\begin{thebibliography}{1}
\bibitem{Chen} Chen J and Davidson J H, Plasma chemistry and plasma processing {\textbf 22} No~4, (2002) 495-522
\bibitem{Aleksandrov} Aleksandrov N L, Bazelyan E M, Carpenter Jr R B, Drabkin M M and Raizer Y P, J. Phys. D: Appl.
Phys. {\textbf 34}, (2001) 3256-3266
\bibitem{Sigmond} Sigmond R S and Lagstad I H, High Temp. Chem. Processes {\textbf 2}, (1993) 221-229
\bibitem{Schlitz} Schlitz D J and Garimella S V, ASME National Heat Transfer Conference, July 2004, (2004) (in
review)
\bibitem{Peek} Peek F W, {\textit Dielectric Phenomena in high voltage engineering} (Mc Graw-Hill Book Company Inc.,
New York 1915 )
\bibitem{Chen2} Chen J and Davidson J H, Plasma chemistry and plasma processing {\textbf 22} No~4, (2002) 199-224
\bibitem{Lowke} Lowke J J and Alessandro F D, J. Phys. D: Appl. Phys. {\textbf 36}, (2003) 2673-2682
\bibitem{Bequin} B\'equin Ph, Castor K and Scholten J, Eur. Phys. J. AP {\textbf 22}, (2003) 41-49
\bibitem{McDonald} McDonald J R, Smith W B, Spencer III H W and Sparks L E, Journal of Applied Physics {\textbf 48}
No~6, (1977) 2231-2243
\bibitem{CG} Shewchunk J R, An introduction to the conjugate gradient method without the agonizing pain
(Pittsburgh: School of Computer Science, Carnegie Mellon University,
Pittsburgh PA 15213 USA 1994 )
\bibitem{NLCG} Bjoerk A, Acta Numerica, (2004) 1-51
\bibitem{Jaiswal} Jaiswal v and Thomas M J, J. Phys. D: Appl. Phys. {\textbf 36}, (2003) 3089-3094
\bibitem{Aboelsaad} Aboelsaad M M, Shafai L and Rashwan M M, IEE Proceedings {\textbf 136} Pt.~A No~1, (1989) 33-40
\end{thebibliography}
\end{document}